\documentclass[12pt]{iopart}
\usepackage{iopams}
\usepackage{graphicx}
\usepackage{makecell}

\newcommand{\mS}{{\mathcal S}}
\newcommand{\mE}{{\mathcal E}}

\newcommand{\mI}{{\mathcal I}}
\newcommand{\mG}{{\mathcal G}}

\begin{document}

\title{Deterministic all-versus-nothing proofs of Bell nonlocality based on non-stabilizer states}

\author{Weidong Tang}

\address{School of Mathematics and Statistics, Shaanxi Normal University, Xi'an 710119, China}
\ead{wdtang@snnu.edu.cn}
\vspace{10pt}

\begin{abstract}
The all-versus-nothing proof of Bell nonlocality is a kind of mainstream demonstration of Bell's theorem without inequalities.  Two kinds of such proofs, called the deterministic  all-versus-nothing proof and the probabilistic  all-versus-nothing proof, are both widely investigated.
So far, all previous deterministic all-versus-nothing proofs of Bell nonlocality are constructed based on stabilizer states. To break with this tradition,  some deterministic all-versus-nothing proofs induced from non-stabilizer states are firstly presented in this work. These results not only can greatly enrich the family of the demonstration of Bell nonlocality without inequalities, but also may provide us some useful resources in certain quantum information processing.
\end{abstract}

\vspace{2pc}
\maketitle

\section{Introduction}
As one of the most striking features of quantum mechanics and a very important quantum resource of quantum information processing, nonlocality plays a significant role in quantum communication and computation. Among various kinds of nonlocality, the most common one is the Bell nonlocality\cite{Bell,Bell-Nonlocality-RMP-2014}, which states that any local realistic model\cite{Bell,CHSH} is incompatible with quantum mechanics. To show this quantum feature, a common approach is to test various of Bell inequalities\cite{Bell,CHSH}, but this approach  can only reveal Bell nonlocality in a probabilistic manner.  By contrast,  another significant approach of demonstrating Bell nonlocality is to construct some logical contradictions. The corresponding proofs  are usually called ``Bell's theorem without inequalities"\cite{GHZ1989,GHZ1990} or ``nonlocality without inequalities"\cite{Hardy92,Hardy93}, which can either be deterministic or probabilistic. The Greenberger-Horne-Zeilinger(GHZ) paradox\cite{GHZ1989,GHZ1990} and the Hardy's paradox\cite{Hardy92,Hardy93} are two typical examples, which belong to two different kinds of all-versus-nothing(AVN) proofs\cite{all-vs-nothing} for Bell nonlocality, respectively.
Precisely, the former provides a  deterministic (or ``always-type") all-versus-nothing (DAVN) proof (which can rule out local realism with a success probability of $100\%$, i.e., it works for each run of the experiment, and thus sometimes also referred to as ``Bell nonlocality without probabilities"\cite{Mermin94}), while the latter provides a probabilistic (or ``sometimes-type") AVN proof (in which the contradiction can only be obtained for some runs of the experiment, and thus the success probability of ruling out local realism is less than $100\%$)\cite{Always-Sometimes-AVN-Cabello-2002}. Apart from  the GHZ paradox and the Hardy's paradox, many generalized versions from them also provide us very wonderful  AVN proofs of Bell nonlocality, such as Cabello's four-qubit AVN proof based on two maximally entangled states\cite{Cabello2001}, the GHZ-type AVN proofs for mixed states\cite{Mix-GHZ-Ghirardi-2006,Mix-GHZ-RCL-2015}, and the multisetting GHZ-type AVN proofs\cite{Multi-Setting-GHZ-TangWeidong2017}. Moreover, some of them have even been tested experimentally\cite{GHZ-exp-Pan-1999,GHZ-exp-Pan-2000,AVN-EXP-Cabello-Cinelli-2005,GHZ-exp-Suzuen-2017}.

It is noted that the quantum states in the DAVN proofs of Bell nonlocality are far more demanding than those in the probabilistic ones. Various GHZ paradoxes are the most common  DAVN proofs, and the involved states are entangled stabilizer states (e.g. GHZ states). In such paradoxes, the systems are usually assumed to be described by local hidden variable (LHV) models, so that value assignments to the local observables (within the involved stabilizers) can be applied. Then one can always obtained a group of value assignment relations
which cannot hold simultaneously, giving rise to  a contradiction. Another representative DAVN proof is Cabello's four-qubit AVN proof\cite{Cabello2001}, wherein the system consisting of two Bell states (note that the whole system is still a stabilizer state) can induce a total of eight independent Hardy-like proofs of Bell nonlocality (each of them can rule out the LHV model with a probability of 12.5$\%$). Combining them together, one can get the desired DAVN proof. In fact,  all of the reported DAVN proofs of Bell nonlocality so far are constructed based on stabilizer states. Then a natural question is: can one construct DAVN proofs of Bell nonlocality based on non-stabilizer states? Since the answer of this question may bring us some new understandings for more refined structures of multi-partite Bell non-locality, to address that is necessary.

On the other hand, it is known that DAVN proofs of Bell nonlocality (e.g. the GHZ paradox) have many important applications in quantum information processing, such as reducing communication complexity\cite{communication-complexity-original,communication-complexity-RMP},
multiparty quantum key distribution\cite{GHZ-key-distribution-Fuyao-2015}, and quantum games\cite{Neighbor-Game-Almeida-2010}. 
It is also known that every property of quantum mechanics not present in classical physics could give rise to an operational advantage\cite{Matera2016,Hillery2016,Theurer2017}. Therefore, if DAVN proofs derived from non-stabilizer states  exist, they can provide
us some useful resources in certain quantum information processing as well. In view of this, the exploration on such unconventional DAVN proofs
is of great value.

\section{A useful tool: the Hardy-like quantum pigeonhole paradox}

To construct a DAVN proof of Bell nonlocality, one should first choose a suitable quantum state. It is worthy to notice that some quantum states can induce more than one probabilistic AVN proof (Hardy-like proof). Also note that combining all such probabilistic AVN proofs (induced from the same quantum state) together may greatly improve the success probability of ruling out local realism\cite{Cabello2001,WeidongTang-HLQP-Paradox}. As long as this success probability can be improved to $100\%$,  a DAVN proof can be produced. In view of this,  the projected-coloring graph(PCG) state proposed very recently\cite{WeidongTang-HLQP-Paradox} seems be a suitable candidate. More precisely, if enough probabilistic AVN proofs can be induced from a PCG state, combining them together will give rise to a DAVN proof. Clearly,  whether the DAVN proof is an unconventional or a conventional one relies on whether the given PCG state is a non-stabilizer state or a stabilizer state.

First, let us give the definition of a PCG state.  For convenience, here we slightly generalize the $n$-qubit PCG state referred to in \cite{WeidongTang-HLQP-Paradox} to the following form:
\begin{equation}\label{PCGS}
    |\Psi\rangle=\frac{1}{\sqrt{|\mI|}}\sum_{i\in\mI}\theta_i|\vec{0}\rangle_{\bar\mS_i}|\vec{1}\rangle_{\mS_i},
\end{equation}
where $\mS_i\subset\{1,2,\cdots,n\}$, $\mS_i\cup\bar{\mS}_i=\{1,2,\cdots,n\}$, $0\leq |\mS_i|<n$, $|\mS_{i}\cup\mS_{j}|>\max\{|\mS_{i}|,|\mS_{j}|\}$ ($i\neq j$), and the coefficients $\theta_{i}\in\{1,-1\}$. Besides, $|\vec{1}\rangle_{\mS_i}\equiv\otimes_{k\in\mS_i}|1\rangle_k$, and $|\vec{0}\rangle_{\bar{\mS}_i}\equiv\otimes_{k\in\bar{\mS}_i}|0\rangle_k$. Moreover, $\mI$ is the index set used for describing a group of specific subsets of $\{1,2,\cdots,n\}$. Note that the original PCG state needs to contain the component $|\vec{0}\rangle_{\{1,2,\cdots,n\}}$, while for the generalized one, such a requirement is not necessary.

The PCG state plays a central role in the construction of a kind of probabilistic AVN proof called the Hardy-like quantum pigeonhole (HLQP) paradox\cite{WeidongTang-HLQP-Paradox}, which can rule out the LHV model\cite{Bell,CHSH} by a logical contradiction from classical pigeonhole principle. Here we shall briefly show this quantum effect by resorting to a simple example.

Let us consider the three-qubit PCG state
\begin{equation}\label{3qubit-PHS}
|\Psi_3\rangle=\frac{1}{2}(|000\rangle-|011\rangle-|101\rangle-|110\rangle).
\end{equation}
One can also check that it is essentially a GHZ state\cite{WeidongTang-HLQP-Paradox}.
Let $|+\rangle=(|0\rangle+|1\rangle)/\sqrt{2}$ and $|-\rangle=(|0\rangle-|1\rangle)/\sqrt{2}$ be two ``boxes" and each qubit be a ``pigeon". Equivalently, if the qubit $k$ stays in box $|+\rangle$, then $X_k=1$; otherwise $X_k=-1$.
Then one can get the following properties (sometimes also referred to as Hardy-like conditions):
\numparts\label{P0}
\begin{eqnarray}
  P(X_2X_3=-1|Z_1=1)&= 1,\label{P0-1}\\
  P(X_1X_3=-1|Z_2=1)&= 1,\label{P0-2}\\
  P(X_1X_2=-1|Z_3=1)&= 1,\label{P0-3}\\
  P(Z_1=1,Z_2=1,Z_3=1) &= 0.25.\label{P0-4}
\end{eqnarray}
\endnumparts
Here $P(X_2X_3=-1|Z_1=1)$, for example,
stands for the conditional probability that $X_2$,$X_3$ are measured and their outcomes satisfy $X_2X_3=-1$  given that the result of $Z_1=1$. Besides, $P(Z_1=1,Z_2=1,Z_3=1)$ is the joint probability of obtaining the results $Z_1=1,~Z_2=1$, and $Z_3=1$.

By invoking the properties of \eref{P0-1} to \eref{P0-4}, one can construct a three-qubit HLQP paradox. To be specific, let three qubits of $|\Phi_3\rangle$ be distributed in different places (hereafter all the measurements are limited to spacelike separated measurements). Consider a run of the experiment that  $Z_1,Z_2$ and $Z_3$ are measured and the results $Z_1=1,Z_2=1$ and $Z_3=1$ are obtained. Assume that $|\Psi_3\rangle$ can be modeled by a LHV theory, namely, for example, the outcome (a predefined value) of measuring $Z_1$ is independent of the choice of the measurements performed on the other qubits. As a consequence, if $X_2$ and $X_3$ were measured in this run, their outcomes should satisfy $X_2X_3=-1$ (pigeon 2 and pigeon 3 are in different boxes). Likewise, similar constraints $X_1X_3=-1$ and $X_1X_2=-1$ must be satisfied in the same run as well. It follows that any pair of the ``pigeons" are staying in different ``boxes", a contradiction by classical pigeonhole principle. Then one can get a three-qubit HLQP paradox.  This paradox shows that any realistic interpretation of quantum mechanics must be nonlocal. For more complicated examples, see \cite{WeidongTang-HLQP-Paradox}.

It is noted that only one type of HLQP paradox (associated with the measurement results $Z_1=Z_2=\cdots=Z_n=1$) was discussed in \cite{WeidongTang-HLQP-Paradox}. In fact, other different types of HLQP paradoxes may also be induced from a given PCG state. For example,
apart from the group of Hardy-like conditions referred to in equation \eref{P0}, different results given by the measurements of $Z_1,Z_2$ and $Z_3$ can induce another three groups of Hardy-like conditions, and besides, each of them can produce a HLQP paradox.  In other words, $|\Psi_3\rangle$ can induce a total of four probabilistic AVN proofs (HLQP paradoxes). Moreover,  one can give a DAVN proof by combining these four  HLQP paradoxes together. However, such a DAVN proof is still induced from a stabilizer state rather than from a non-stabilizer one, which is similar to the DAVN proof of Bell's theorem  proposed in \cite{Cabello2001}. For detailed discussion, see \ref{app-a}.

\section{The pictorial representation of a group of Hardy-like conditions}

Since a HLQP paradox is associated with a group of Hardy-like conditions,  as long as a graphical structure can faithfully represent this group of Hardy-like conditions, it can give a pictorial representation for  the HLQP paradox. Based on this thought,
any original HLQP paradox can be pictorially represented by a mathematical structure called the PCG\cite{WeidongTang-HLQP-Paradox}. Technically, similar representations can also be generalized other types of HLQP paradoxes, and for simplicity, such generalized  representations are still referred to as PCGs. Next we will show how to use a PCG to represent a group of Hardy-like conditions.

Consider the following group of constraints induced from the aforementioned $n$-qubit PCG state $|\Psi\rangle$,
\begin{eqnarray}\label{generalized-PCG}
P(\prod_{k\in\mE_1}X_k=\alpha_1|\vec{Z}_{\bar{\mE}_1}=\vec{m}_{\bar{\mE}_1})&=1,\cr
P(\prod_{k\in\mE_2}X_k=\alpha_2|\vec{Z}_{\bar{\mE}_2}=\vec{m}_{\bar{\mE}_2})&=1,\cr
\vdots~~~~~~~~~&\cr
P(\prod_{k\in\mE_r}X_k=\alpha_r|\vec{Z}_{\bar{\mE}_r}=\vec{m}_{\bar{\mE}_r})&=1,\cr
P(Z_1=m_1,Z_2=m_2,\cdots,Z_n=m_n)&=\frac{1}{|\mI|},
\end{eqnarray}
where $\mE_j\subset\{1,2,\cdots,n\}$ and $\alpha_j\in\{1,-1\}$ $(j=1,2,\cdots,r)$. Besides, $\vec{Z}_{\bar{\mE}_j}=\vec{m}_{\bar{\mE}_j}$ stands for $Z_{t_1(j)}=m_{t_1(j)},Z_{t_2(j)}=m_{t_2(j)},\cdots,Z_{t_{s_j}(j)}=m_{t_{s_j}(j)}$, where $\bar{\mE}_j=\{t_1(j),t_2(j),\cdots,t_{s_j}(j)\}$, and $\mE_j\cup\bar{\mE}_j=\{1,2,\cdots,n\}$. For convenience, we would still call these constraints a group of Hardy-like conditions regardless of whether they can induce a HLQP paradox or not. Then the PCG corresponding to this group of  Hardy-like conditions can be defined as follows:

(i) Each vertex is represented either by $\bullet$ or by $\circ$. To be specific, if $m_i=1$, we  choose $\bullet$ to represent the $i$-th vertex; otherwise, we use $\circ$ to represent this vertex.

(ii)Each edge is represented by a closed green or red curve circulating at least two vertices. Note that the $j$-th edge $\mE_j$ is used for
describing the (conditional) constraint $\prod_{k\in\mE_j}X_k=\alpha_j$ associated with the $j$-th conditional probability relation $P(\prod_{k\in\mE_j}X_k=\alpha_j|\vec{Z}_{\bar{\mE}_j}=\vec{m}_{\bar{\mE}_j})=1$ in equation \eref{generalized-PCG}. To be specific, if $\alpha_j=-1$, the edge $\mE_j$ is represented by a closed red curve which circulates all the vertices in  $\mE_j$; otherwise, $\mE_j$ is represented by a closed green curve.
\begin{figure}[t]
  \centering
\includegraphics[scale=1]{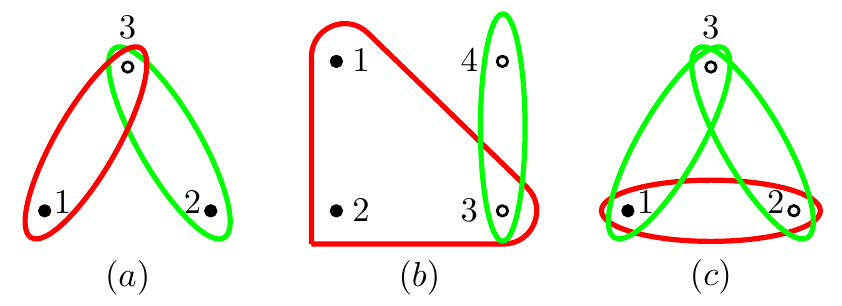} \caption{\label{fig0} Examples of PCGs. (a) The PCG associated with the Hardy-like conditions: $P(Z_1=1,Z_2=1,Z_3=-1)=\frac{1}{3},~P(X_2X_3=1|Z_1=1)=1,~P(X_1X_3=-1|Z_2=1)=1$, which are induced from the PCG state $(|001\rangle+|010\rangle-|100\rangle)/\sqrt{3}$. (b) The PCG corresponds to the Hardy-like conditions: $P(Z_1=1,Z_2=1,Z_3=-1,Z_4=-1)=\frac{1}{3}, ~P(X_1X_2X_3=-1|Z_4=-1)=1,~P(X_3X_4=1|Z_1=1,Z_2=1)=1$,  where the PCG state is $(|0000\rangle-|1101\rangle+|0011\rangle)/\sqrt{3}$. (c) An un-colorable loop PCG as a pictorial representation of the Hardy-like conditions $P(Z_1=1,Z_2=-1,Z_3=-1)=\frac{1}{4}, ~P(X_2X_3=1|Z_1=1)=1,~P(X_1X_3=1|Z_2=-1)=1,~P(X_1X_2=-1|Z_3=-1)=1$. The corresponding PCG state is $(|000\rangle+|011\rangle-|101\rangle+|110\rangle)/2$. }
\end{figure}

{\it Example 1.} Consider the PCG state $(|001\rangle+|010\rangle-|100\rangle)/\sqrt{3}$. It can induce a group of Hardy-like conditions:  $P(Z_1=1,Z_2=1,Z_3=-1)=\frac{1}{3}, ~P(X_2X_3=1|Z_1=1)=1,~P(X_1X_3=-1|Z_2=1)=1$. As $m_1=1,~m_2=1$, and $m_3=-1$, Both of the vertices 1 and 2 in the PCG should be represented by $\bullet$, and besides, the vertex 3 should be represented by $\circ$. Moreover, from the relation $P(X_2X_3=1|Z_1=1)=1$, one can conclude that the edge $\{2,3\}$ should be colored with green. Likewise, $P(X_1X_3=-1|Z_2=1)=1$ indicates that the edge $\{1,3\}$ is colored with red. Then the final PCG can be obtained,  see \fref{fig0}(a). \hfill$\blacksquare$

{\it Example 2.} The PCG state is $(|0000\rangle-|1101\rangle+|0011\rangle)/\sqrt{3}$, and the Hardy-like conditions are $P(Z_1=1,Z_2=1,Z_3=-1,Z_4=-1)=\frac{1}{3}, ~P(X_1X_2X_3=-1|Z_4=-1)=1,~P(X_3X_4=1|Z_1=1,Z_2=1)=1$. From $m_1=m_2=1,m_3=m_4-1$, we know that the vertices 1 and 2 are represented by $\bullet$, and the vertices 3 and 4 are represented by $\circ$. Moreover, the relation $P(X_1X_2X_3=-1|Z_4=-1)=1$ gives rise to a red edge $\{1,2,3\}$, while  $P(X_3X_4=1|Z_1=1,Z_2=1)=1$ indicates that the edge $\{3,4\}$ is green. For the final PCG,
see \fref{fig0}(b). \hfill$\blacksquare$

Clearly, the Hardy-like conditions in the above two examples cannot give rise to HLQP paradoxes. This also indicates that not all the PCGs can be associated to HLQP paradoxes.   Similar to  the discussion in \cite{WeidongTang-HLQP-Paradox}, to determine whether a given PCG is a pictorial representation of a HLQP paradox, one can consider such an equivalent vertex-coloring problem:
For an $n$-qubit PCG $\mG$ with vertex-set $\{1,2,\cdots,n\}$ and edge set $\{\mE_1,\mE_2,\cdots,\mE_r\}$, check whether there exists a consistent coloring scheme for all the vertices, wherein the vertex-coloring rules are described as follows.

(a) Each vertex can be colored with either green or red. If the $k$-th vertex is colored with red, its coloring value is defined as $c_k=-1$; otherwise $c_k=1$.

(b) If the $i$-th edge $\mE_i$ is red,  the corresponding weight is defined as $W(\mE_i)=-1$; otherwise $W(\mE_i)=1$.

(c) If there exist at least one coloring scheme such that $\prod_{k\in\mE_i}c_k=W(\mE_i)$ holds for any $\mE_i\in\{\mE_1,\mE_2,\cdots,\mE_r\}$, the PCG $\mG$ is colorable; otherwise, $\mG$ is un-colorable.

Note that the color of the edge $\mE_j$ is determined by the constraint $\prod_{k\in\mE_j}X_k=\alpha_j$. Besides, it is known that if the group of the constraints $\prod_{k\in\mE_1}X_k=\alpha_1$, $\prod_{k\in\mE_2}X_k=\alpha_2$, $\cdots$, $\prod_{k\in\mE_r}X_k=\alpha_r$ cannot hold  simultaneously, a HLQP paradox can be constructed. Then according to the above vertex-coloring rules,
only un-colorable PCGs can be associated to HLQP paradoxes.

Furthermore, if all of the $|\mI|$ PCGs induced from the PCG state $|\Psi\rangle$ are un-colorable PCGs, a DAVN proof can be derived, see a minimal example in \ref{app-b}. Note that sometimes pictorial representations (by PCGs) can be used for quick identification of the desired DAVN proofs.
For example, a loop PCG (in which each edge only connects two vertices and each vertex is shared by two adjacent edges) with an odd number of edges colored with red is an un-colorable PCG. In fact, \fref{fig0}(c) shows us one of the simplest un-colorable loop PCGs, wherein only one edge is colored with red. One can also check that as long as the PCG contains a loop substructure in which there is an odd number of red edges, it is an un-colorable PCG. Apart from that, it is obvious that similar techniques apply to the PCGs containing other un-colorable substructures as well.

\section{Main results}

In this section, we shall construct some typical DAVN proofs of Bell nonlocality based on non-stabilizer states.

\noindent\begin{center}
\begin{figure}[t]
  \centering
\includegraphics[scale=1]{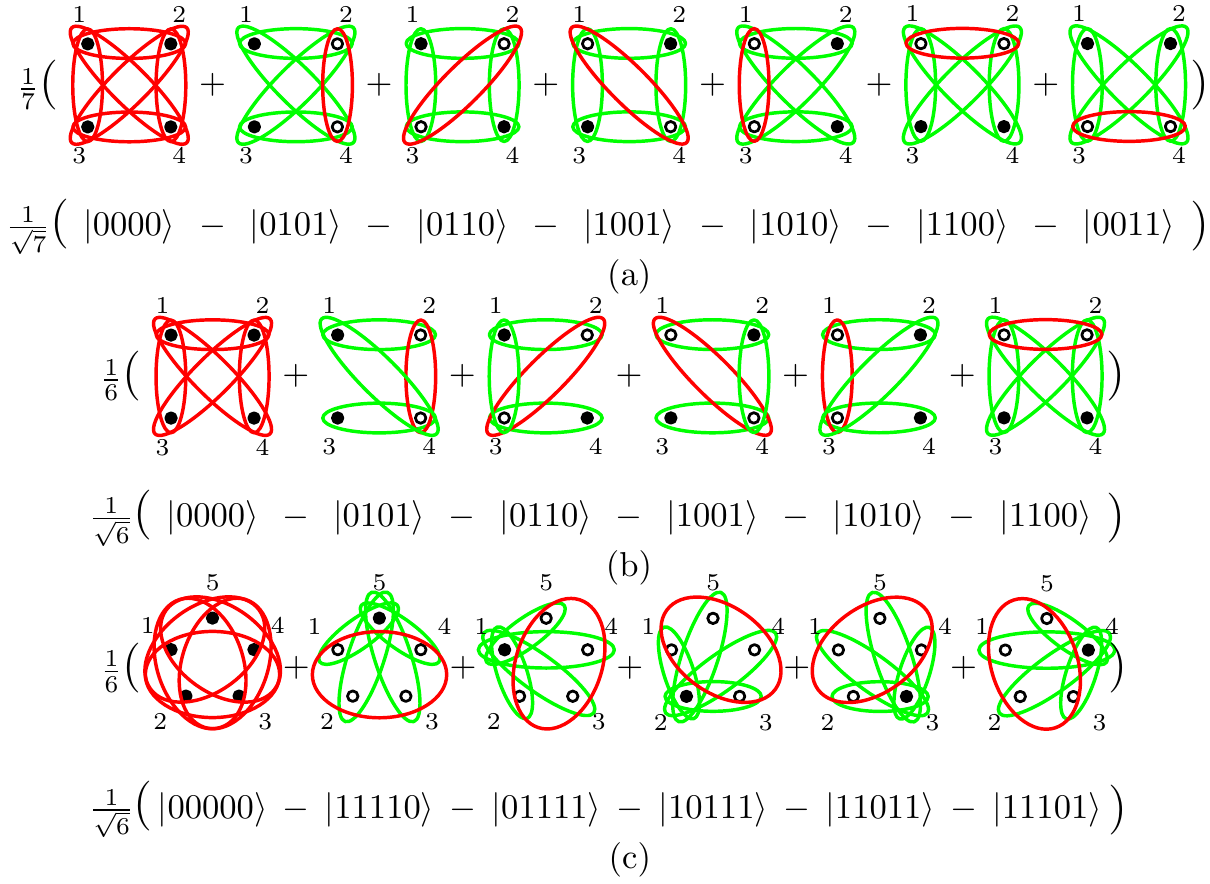} \caption{\label{fig2} Pictorial representations for three DAVN proofs of Bell nonlocality from non-stabilizer states, in which each PCG is an un-colorable PCG, and it corresponds to a HLQP paradox.
 Besides, the coefficient of each PCG, such as $\frac{1}{7}$ or $\frac{1}{6}$,  describes the proportion of each HLQP paradox in the corresponding DAVN proof.}
\end{figure}
\end{center}

Our first DAVN of Bell nonlocality from non-stabilizer states is induced from the following symmetric four-qubit PCG state
\begin{equation}\label{5qubit-type-A}
|\Phi_4\rangle=\frac{1}{\sqrt{7}}(|0000\rangle-|S(2,2)\rangle),
\end{equation}
where $|S(2,2)\rangle$ is the sum of all permutations of $\underbrace{|0\rangle|0\rangle}_{2}\underbrace{|1\rangle|1\rangle}_2$, i.e., $|S(2,2)\rangle=|0011\rangle+|0101\rangle+|0110\rangle+|1001\rangle+|1010\rangle+|1100\rangle$.

One can check whether $|\Phi_4\rangle$ is a stabilizer state by calculating the reduced density operator on  each qubit. In fact, as long as there exists at least one reduced density operator on single qubit not equal to $\frac{I}{2}$, the state is a non-stabilizer state.

To see that,  let us consider the stabilizer states which are fully entangled (since correlations between disentangled subsystems cannot reveal Bell nonlocality). Assume that $|S\rangle$ is such an $n$-qubit stabilizer state, and the $n$ independent stabilizers are
$g_1,g_2,\cdots,g_n$.  Then one have $|S\rangle\langle S|=\prod_{i=1}^n\frac{I+g_i}{2}$. Note that each $g_i$ is a tensor product of Pauli operators and a variable number of identity operators (up to a local unitary transformation), and such a tensor product requires at least two Pauli operators from different qubits (otherwise it would contradict with the fact that the stabilizer state is fully entangled). Moreover, the product of $k$ ($2\leq k\leq n$) different stabilizers is still a stabilizer, and in final form the this product at least two Pauli operators from different qubits are included. Thus, one can get a set of constraints for the partial traces: $\tr_{V\backslash k}(g_i)=0,\tr_{V\backslash k}(g_ig_j)=0,\tr_{V\backslash k}(g_ig_jg_k)=0,\cdots$, where $V=\{1,2,\cdots,n\}$, $V\backslash k\equiv\{1,2,\cdots,k-1,k+1,\cdots,n\}$, and $i\neq j\neq k\neq \cdots$. Then the reduced density operator on the $i$-th qubit is
$\rho_i=\tr_{V\backslash i}|S\rangle\langle S|=\frac{I}{2}$. This also indicates that if a reduced density operator on single qubit of a fully entangled state is not equal to $\frac{I}{2}$, this entangled state state must be a non-stabilizer state.

A straightforward calculation shows that the reduced density operator $\rho_i=(4|0\rangle\langle0|+3|1\rangle\langle1|)/7\neq \frac{I}{2}$ ($i=1,2,3,4$), indicating that $|\Phi_4\rangle$ is a non-stabilizer state.

It turns out that seven groups of Hardy-like conditions can be derived from $|\Phi_4\rangle$.

(I) The first group of Hardy-like conditions can be written as
\begin{equation}\label{1class-J}
  P(Z_1=1,Z_2=1,Z_3=1,Z_4=1) = \frac{1}{7},
\end{equation}
and
\begin{equation}\label{1class-C}
  P(X_kX_l=-1|Z_i=Z_j=1)= 1,
\end{equation}
where $i\neq j\neq k\neq l\in\{1,2,3,4\}$.
Note that equation \eref{1class-C} contains 6 conditional probability relations.

To construct a HLQP paradox, let us
consider  a run of the experiment that  $Z_1,Z_2,Z_3$ and $Z_4$ are measured and the results $Z_1=1,Z_2=1,Z_3=1$ and $Z_4=1$ are obtained (happens with a probability of $\frac{1}{7})$.
Assume that $|\Phi_4\rangle$ admits a LHV model. Similar to the argument in the aforementioned three-qubit HLQP paradox, since one have $Z_3=Z_4=1$, one can infer from  equation \eref{1class-C} that if $X_1,X_2$ were measured, their results must satisfy $X_1X_2=-1$. Likewise, five other constraints can be obtained as well.
Precisely, according to the LHV model, if $X_1,X_2,X_3$ and $X_4$ were measured in this run, one would get $X_1X_2=X_1X_3=X_1X_4=X_2X_3=X_2X_4=X_3X_4=-1$.  This indicates that when four pigeons are put into two boxes in this run, any pair of the pigeons cannot stay in the same box, a contradiction by classical pigeonhole principle. Then a HLQP paradox can be obtained.

(II) The other six groups of Hardy-like conditions have the same structure (up to a permutation): For any given $i\neq j\neq k\neq l\in\{1,2,3,4\}$,
\begin{equation}\label{2class-J}
  P(Z_i=-1,Z_j=-1,Z_k=1,Z_l=1) = \frac{1}{7},
\end{equation}
and
\numparts\label{2class-C}
\begin{eqnarray}
  &P(X_jX_l=1|Z_i=-1,Z_k=1)= 1,\label{2class-C-1}\\
  &P(X_jX_k=1|Z_i=-1,Z_l=1)= 1,\label{2class-C-2}\\
  &P(X_iX_l=1|Z_j=-1,Z_k=1)= 1,\label{2class-C-3}\\
  &P(X_iX_k=1|Z_j=-1,Z_l=1)= 1,\label{2class-C-4}\\
  &P(X_iX_j=-1|Z_k=1,Z_l=1)= 1.\label{2class-C-5}
\end{eqnarray}
\endnumparts

Consider  a run of the experiment that  $Z_i,Z_j,Z_k$ and $Z_l$ are measured and  the results $Z_i=-1,Z_j=-1,Z_k=1$ and $Z_l=1$ (associated with one of the components in $|S(2,2)\rangle$) are obtained. As long as the quantum state $|\Phi_4\rangle$ admits
a LHV model, one can always conclude that if $X_i,X_j,X_k$ and $X_l$ were measured in this run, their values must satisfy $X_iX_j=-1, X_iX_k=X_iX_l=X_jX_k=X_jX_l=1$, which also violate the pigeonhole counting principle. Then one can get six such
HLQP paradoxes.

To sum up, (I) and (II) can give a total of seven HLQP paradoxes.

Since in any run of the experiment, the measurements of $Z_1,Z_2,Z_3$ and $Z_4$ can only give seven groups of results, and for every group of the results, one can always invoke one of the proofs of seven HLQP paradoxes referred in (I) and (II) to exclude the local realistic description of the quantum system. To be specific, seven groups of results $\{Z_1=Z_2=Z_3=Z_4=1\}$ and $\{Z_i=-1,Z_j=-1,Z_k=1,Z_l=1\}$ ($i\neq j\neq k\neq l\in\{1,2,3,4\}$) can give rise to seven value assignment  contradictions, i.e., one is
$\{X_1X_2=X_1X_3=X_1X_4=X_2X_3=X_2X_4=X_3X_4=-1\}$ and the other six can be described by $\{X_iX_j=-1,X_iX_k=X_iX_l=X_jX_k=X_jX_l=1\}$ ($i\neq j\neq k\neq l\in\{1,2,3,4\}$). Therefore, combining the seven HLQP paradoxes together, one can get a DAVN proof (from a non-stabilizer state). For its pictorial representation, see \fref{fig2}(a).

Inspired by this example,  a family of  DAVN proofs based on $n$-qubit ($n\geq4$) non-stabilizer states can be analytically constructed.
Detailed discussion is shown in \ref{app-c}.

In addition,  another four-qubit example of DAVN proof for Bell nonlocality can be constructed base on the four-qubit non-stabilizer state $|\Phi_4^{\prime}\rangle=(|0000\rangle-|0101\rangle-|0110\rangle-|1001\rangle-|1010\rangle-|1100\rangle)/\sqrt{6}$. One can check that by \fref{fig2}(b). Moreover,  some states which are locally unitary equivalent to $|\Phi_4^{\prime}\rangle$ can also induce similar DAVN proofs. A typical example is shown in  \ref{app-d}.

Next, we shall present a five-qubit DAVN proof of Bell nonlocality based on the following following PCG state
\begin{equation}\label{5qubit-type-B}
|\phi_5\rangle=\frac{1}{\sqrt{6}}(|00000\rangle-|S(1,4)\rangle),
\end{equation}
where $|S(1,4)\rangle=|01111\rangle+|10111\rangle+|11011\rangle+|11101\rangle+|11110\rangle$.

Notice that the reduced density operator on the $i$-th qubit is $\rho_i=(|0\rangle\langle0|+2|1\rangle\langle1|)/3\neq \frac{I}{2}$ ($i=1,2,3,4,5$), and thus $|\phi_5\rangle$ is also a non-stabilizer state.
One can check that $|\phi_5\rangle$ can induce six groups of Hardy-like conditions.

(I$^\prime$)  The first group of Hardy-like conditions can be described as follows:
\begin{equation}\label{5-TB-N-J}
  P(Z_1=1,Z_2=1,Z_3=1,Z_4=1,Z_5=1) = \frac{1}{6},
\end{equation}
and
\begin{equation}\label{5-TB-N-C}
  P(X_jX_kX_lX_m=-1|Z_i=1)= 1,
\end{equation}
where $i\neq j\neq k\neq l\neq m\in\{1,2,3,4,5\}$.
Note that equation \eref{5-TB-N-C} contains  five conditional probability relations.

Consider a run of the experiment that $Z_1,Z_2,Z_3,Z_4,Z_5$ are measured and their outcomes $Z_1=1,Z_2=1,Z_3=1,Z_4=1,Z_5=1$ are obtained (with a probability of $\frac{1}{6}$). Assume that $|\phi_5\rangle$ admits a LHV model. According to equation \eref{5-TB-N-C}, since we have $Z_1=1$, if $X_2,X_3,X_4,X_5$ were measured in this run, one can infer that their results need to satisfy $X_2X_3X_4X_5=1$. Similarly, other four constraints can be derived as well, i.e., $X_1X_3X_4X_5=X_1X_2X_4X_5=X_1X_2X_3X_5=X_1X_2X_3X_4=-1$. According to classical  pigeonhole principle,  all the five constraints cannot hold simultaneously. Then one can get a HLQP paradox.

(II$^\prime$) The other five  Hardy-like conditions belong to the same type (up to a permutation), and each of them can be specified as
\begin{equation}\label{5-TB-5-J}
  P(Z_i=1,Z_j=-1,Z_k=-1,Z_l=-1,Z_m=-1) = \frac{1}{6},
\end{equation}
and
\numparts\label{5-TB-5-C}
\begin{eqnarray}
  &P(X_jX_kX_lX_m=-1|Z_i=1)= 1,\label{5-TB-5-C-1}\\
  &P(X_iX_j=1|Z_k=Z_l=Z_m=-1)=1,\label{5-TB-5-C-2}\\
  &P(X_iX_k=1|Z_l=Z_m=Z_j=-1)=1,\label{5-TB-5-C-3}\\
  &P(X_iX_l=1|Z_m=Z_j=Z_k=-1)=1,\label{5-TB-5-C-4}\\
  &P(X_iX_m=1|Z_j=Z_k=Z_l=-1)=1,\label{5-TB-5-C-5}
\end{eqnarray}
\endnumparts
where $i\neq j\neq k\neq l\neq m\in\{1,2,3,4,5\}$.

Likewise, once $Z_i=1,Z_j=Z_k=Z_l=Z_m=-1$ are obtained in some run of the experiment, one can infer that if $X_1,X_2,X_3,X_4,X_5$ were measured in this run, their results must satisfy $X_jX_kX_lX_m=-1,X_iX_j=X_iX_k=X_iX_l=X_iX_m=1$ according to local realism.
Clearly, such constraints also contradict with classical pigeonhole Principle. Therefore, five HLQP paradoxes can be constructed from this type of Hardy-like conditions.

Notice that measuring $Z_1,Z_2,Z_3,Z_4,Z_5$ of $|\phi_5\rangle$ can only give six groups of results, and according to (I$^\prime$) and (II$^\prime$), any of them can induce a HLQP paradox. Then combining them together will give rise to a DAVN proof, also see \fref{fig2}(c) for its pictorial representation.

Moreover, by using a analytical construction technique, one can generalize this example to a family of $(2n+3)$-qubit ($n\geq1$) DAVN proofs.
For more details, see \ref{app-e}.

In the end, some remarks are in order.  First,
notice that sometimes permutation symmetries can greatly simplify the description of the involved HLQP paradoxes (some of them belong to the same class). Therefore, choosing a permutation symmetrical system or a roughly permutation symmetrical system (one or a few components removed away from a permutation symmetrical state) may simplify the construction of the DAVN proof.
Second, note that some quantum systems in aforementioned DAVN proofs seem to be ``very close" to stabilizer states. For example, $|\Phi_4\rangle$ is very close to the fully entangled  stabilizer state (a four-qubit GHZ state) $|S_4\rangle=(|\circlearrowleft\circlearrowleft\circlearrowleft\circlearrowleft\rangle
+|\circlearrowright\circlearrowright\circlearrowright\circlearrowright\rangle)/{\sqrt{2}}=(|0000\rangle+|1111\rangle-|S(2,2)\rangle)/{(2\sqrt{2})}$.
This is not curious. In fact, the $4$-qubit  stabilizer state $|S_4\rangle$ can induce a GHZ paradox,  but this paradox is not a genuine $4$-partite GHZ paradox but only a genuine $3$-partite one\cite{Cerf2002}. Also note that this GHZ paradox can be converted to a combination of eight HLQP paradoxes, and moreover,  for each HLQP paradox, the corresponding Hardy-like conditions can provides far more relations than it really needs. Thus even if one of the terms in $|S_4\rangle$ were missing, each group of new Hardy-like conditions induced from the remaining ones (e.g. the remaining terms form the state $|\Phi_4\rangle$) may still give rise to a HLQP paradox.
In view of these, in a practical construction of the unconventional DAVN proof, we prefer to choose the non-stabilizer state which has permutation symmetries (or roughly permutation symmetries) and is very close to some stabilizer state.

\section{Conclusion}

To summarize,  some four- and five-qubit DAVN proofs of Bell nonlocality based on non-stabilizer states have been presented, opening a new chapter in the study of DAVN proofs of Bell nonlocality,  as previous ones were always induced from stabilizer states. Besides, such unconventional  proofs can also be  generalized to the scenarios with more qubits analytically, and as a consequence, one can get several families of scalable DAVN proofs of Bell nonlocality. Our results can not only add many new members to the family of AVN proofs, but also help us to get a better understanding of more refined structures for multi-party Bell nonlocality.  However, due to the dramatic growth of calculation, how to derive the DAVN proof of Bell nonlocality from a qudit non-stabilizer state would be a huge challenge. Until very recently, we found a non-trivial example with $d=4$, which was reported in another work.

\ack

We thank K. Han, W. Du  and D. Zhou for helpful discussions. This work was supported by Natural Science Basic Research Plan in Shaanxi Province of China (Program No. 2023-JC-YB-035).

\appendix

\section{A  DAVN proof for Bell nonlocality from the three-qubit GHZ state}\label{app-a}
In fact, the three qubit GHZ state $|\Psi_3\rangle=(|000\rangle-|011\rangle-|101\rangle-|110\rangle)/2$ can give a total of four groups of
Hardy-like conditions, See \tref{TB1}-(I-IV). Clearly, each of them can produce a HLQP paradox. Then $|\Psi_3\rangle$ can induce a total of four HLQP paradoxes.

\begin{table*}
\caption{Four groups of Hardy-like conditions  induced from $|\Psi_3\rangle$.} \label{TB1}
\begin{center}
\lineup\begin{tabular}{ccc}
\br
  I & \thead{$P(X_2X_3=-1|Z_1=1) = 1$,\\ $P(X_1X_3=-1|Z_2=1) = 1$,} & \thead{$P(X_1X_2=-1|Z_3=1) = 1$,\\ $P(Z_1=1,Z_2=1,Z_3=1) = 0.25$} \\
 \mr
  II& \thead{$P(X_2X_3=1|Z_1=-1) = 1$,\\ $P(X_1X_3=1|Z_2=-1) = 1$,} & \thead{$P(X_1X_2=-1|Z_3=1) = 1$,\\ $P(Z_1=-1,Z_2=-1,Z_3=1) = 0.25$}\\
 \mr
  III& \thead{$P(X_2X_3=-1|Z_1=1) = 1$,\\ $P(X_1X_3=1|Z_2=-1) = 1$,} & \thead{$P(X_1X_2=1|Z_3=-1) = 1$,\\ $P(Z_1=1,Z_2=-1,Z_3=-1) = 0.25$}\\
\mr
  IV & \thead{$P(X_2X_3=1|Z_1=-1) = 1$,\\ $P(X_1X_3=-1|Z_2=1) = 1$,} & \thead{$P(X_1X_2=1|Z_3=-1) = 1$,\\ $P(Z_1=-1,Z_2=1,Z_3=-1) = 0.25$}\\
\br
\end{tabular}
\end{center}
\end{table*}

Combining these HLQP paradoxes together, one can get a DAVN proof for Bell nonlocality. To show that, consider any run of the experiment that $Z_1,Z_2$ and $Z_3$ are measured. According to  \tref{TB1}, only four groups of results are possible (each occurs with a probability of $25\%$). If the quantum state  admits
a LHV model,  a contradiction can be obtained from any group of the results. Precisely, $\{Z_1=1,Z_2=1,Z_3=1\},\{Z_1=1,Z_2=-1,Z_3=-1\},\{Z_1=-1,Z_2=1,Z_3=-1\}$ and  $\{Z_1=-1,Z_2=-1,Z_3=1\}$ will give rise to $\{X_2X_3=-1,X_1X_3=-1,X_1X_2=-1\},\{X_2X_3=-1,X_1X_3=1,X_1X_2=1\},\{X_2X_3=1,X_1X_3=-1,X_1X_2=1\}$ and $\{X_2X_3=1,X_1X_3=1,X_1X_2=-1\}$, respectively. Clearly, all these relations violate classical pigeonhole principle.

Note that the GHZ state $|\Psi_3\rangle$ is a stabilizer state, this DAVN proof is still a conventional demonstration of Bell nonlocality without inequalities.

\section{Pictorial representation for a three-qubit DAVN proof for Bell nonlocality}\label{app-b}
\begin{center}
\begin{figure}[h]
\centering
\includegraphics[scale=1]{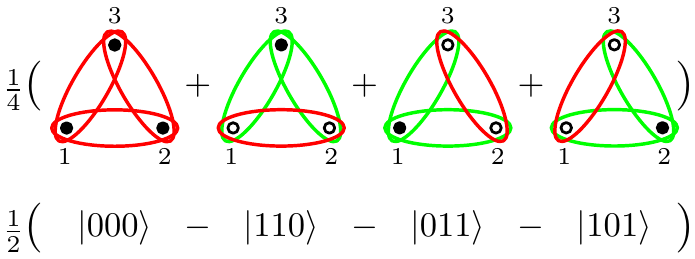} \caption{\label{fig-app-1} Pictorial representation  for the DAVN proof of Bell nonlocality induced from $|\Psi_3\rangle=(|000\rangle-|011\rangle-|101\rangle-|110\rangle)/2$. The four PCGs correspond to type I-IV HLQP paradoxes in \tref{TB1} respectively. The coefficient $1/4$ stands for the proportion of each HLQP paradox in this DAVN proof.}
\end{figure}
\end{center}

Notice that the PCG can give an intuitive pictorial representation for the HLQP paradox. Combining the associated PCGs together, one can get  the pictorial representation for the DAVN proof, see \fref{fig-app-1}.

\section{An example for $n$ ($n\geq4$) qubits}\label{app-c}
Consider  the following $n$-qubit ($n\geq4$) projected-coloring graph state
\begin{equation}\label{n-qubit-PCGS}
|\Phi_{n}\rangle=\frac{1}{\sqrt{C_{n}^2+1}}(|00\cdots0\rangle-|S(n-2,2)\rangle),
\end{equation}
where $|S(n-2,2)\rangle$ is the sum of all permutations of $\underbrace{|0\rangle|0\rangle\cdots|0\rangle}_{n-2}\underbrace{|1\rangle|1\rangle}_2$, namely, $|S(n-2,2)\rangle=|00\cdots011\rangle+|00\cdots101\rangle+|00\cdots110\rangle+\cdots+|11\cdots000\rangle$.

Since $|\Phi_{n}\rangle$ is a fully entangled state, and a straightforward calculation shows that the reduced density operator on the $i$-th qubit is $\rho_i=[(C_{n}^2-n+1)|0\rangle\langle0|+n|1\rangle\langle1|]/(C_{n}^2+1)\neq \frac{I}{2}$ ($i=1,2,\cdots,n$), one can conclude that $|\Phi_n\rangle$ is a non-stabilizer state.

We can get $(C_{n}^2+1)$  groups of Hardy-like conditions (which belongs to two classes) from $|\Phi_{n}\rangle$.

The first class (contains only one group):
\begin{equation}\label{1class-J-odd}
  P(Z_1=1,Z_2=1,\cdots,Z_{n}=1) = \frac{1}{C_{n}^2+1},
\end{equation}
and
\begin{equation}\label{1class-C-odd}
  P(X_{i_1}X_{i_2}=-1|Z_{i_3}=Z_{i_4}=\cdots=Z_{i_{n}}=1)= 1,
\end{equation}
where $i_1\neq i_2\neq \cdots\neq i_{n}\in\{1,2,\cdots,n\}$.
Note that equation \eref{1class-C-odd} contains $C_{n}^2$ such relations.

The second class (contains $C_{n}^2$ groups, only one of them is listed in the following): For any given $i_1\neq i_2\neq \cdots\neq i_{n}\in\{1,2,\cdots,n\}$,
\begin{equation}\label{2class-J-odd}
  P(Z_{i_1}=-1,Z_{i_2}=-1,Z_{i_3}=1,\cdots,Z_{i_{n}}=1) = \frac{1}{C_{n}^2+1},
\end{equation}
and
\numparts\label{2class-C-odd}
\begin{eqnarray}
  &P(X_{i_2}X_{j_{n-2}}=1|Z_{i_1}=-1,Z_{j_1}=Z_{j_2}=\cdots=Z_{j_{n-3}}=1)= 1,\label{2class-C-odd-1}\\
  &P(X_{i_1}X_{j_{n-2}}=1|Z_{i_2}=-1,Z_{j_1}=Z_{j_2}=\cdots=Z_{j_{n-3}}=1)= 1,\label{2class-C-odd-2}\\
  &P(X_{i_1}X_{i_2}=-1|Z_{i_3}=Z_{i_4}=\cdots=Z_{i_{n}}=1)= 1.\label{2class-odd-C-3}
\end{eqnarray}
\numparts
Here $j_1\neq j_2\neq\cdots\neq j_{n-2}\in\{1,2,\cdots,n\}\backslash\{i_1,i_2\}$. Note that
both of equation \eref{2class-C-odd-1} and equation \eref{2class-C-odd-2} contain $n-2$ relations.

Assume that the quantum system admits a LHV model, then one can prove that each group of these Hardy-like conditions conflicts with classical pigeonhole principle. As a consequence, a total of $(C_{n}^2+1)$ HLQP paradoxes can be obtained. Combining them together will produce
an $n$-qubit($n\geq4$) DAVN proof of Bell nonlocality.

\section{Another four-qubit unconventional DAVN proof}\label{app-d}
\begin{center}
\begin{figure*}[t]
  \centering
\includegraphics[scale=1]{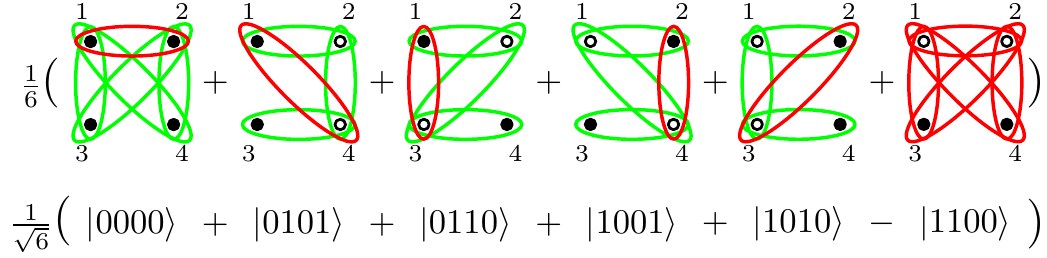} \caption{\label{fig-app} Pictorial representations for the DAVN proof induced from $(|0000\rangle+|0101\rangle+|0110\rangle+|1001\rangle+|1010\rangle-|1100\rangle)/\sqrt{6}$, in which each PCG is a representation of a HLQP paradox.}
\end{figure*}
\end{center}

Consider the state  $|\Phi_4^{\prime\prime}\rangle=(|0000\rangle+|0101\rangle+|0110\rangle+|1001\rangle+|1010\rangle-|1100\rangle)/\sqrt{6}$. Similarly, combining the six HLQP paradoxes (associated with all possible  measurement results of $Z_1,Z_2,Z_3$ and $Z_4$) together, we can construct a DAVN proof of Bell nonlocality. For the pictorial representation, see \fref{fig-app}.

\section{An example for $2n+3$ ($n\geq1$) qubits}\label{app-e}
Consider  the following $(2n+3)$-qubit PCG state
\begin{equation}\label{2n3-qubit-PCGS}
|\phi_{2n+3}\rangle=\frac{1}{\sqrt{2n+4}}(|00\cdots0\rangle-|S(1,2n+2)\rangle),
\end{equation}
where $|S(1,2n+2)\rangle$ is the sum of all permutations of $\underbrace{|0\rangle}_{1}\underbrace{|1\rangle|1\rangle\cdots|1\rangle}_{2n+2}$, namely, $|S(1,2n+2)\rangle=|011\cdots11\rangle+|101\cdots11\rangle+|110\cdots11\rangle+\cdots+|11\cdots110\rangle$.

Note that $|\phi_{2n+3}\rangle$ is a fully entangled state, and one can check that the reduced density operator on the $i$-th qubit is $\rho_i=[|0\rangle\langle0|+(n+1)|1\rangle\langle1|]/(n+2)\neq \frac{I}{2}$ ($i=1,2,\cdots,2n+3$).  Therefore, $|\phi_{2n+3}\rangle$ is a non-stabilizer state.

Here we can get $2n+4$  groups of Hardy-like conditions (which belong to two classes).

The first class (contains only one group):
\begin{equation}\label{1class-J-odd-B}
  P(Z_1=1,Z_2=1,\cdots,Z_{2n+3}=1) = \frac{1}{2n+4},
\end{equation}
and
\begin{equation}\label{1class-C-odd-B}
  P(X_{j_1}X_{j_2}\cdots X_{j_{2n+2}}=-1|Z_i=1)= 1,
\end{equation}
where  $i\neq j_1\neq j_2\neq \cdots\neq j_{2n+2}\in\{1,2,\cdots,2n+3\}$.
Note that equation \eref{1class-C-odd-B} contains $2n+3$ such relations.

The second class (contains $2n+3$ groups, and only one of them is listed in the following):  The $i$-th group of such Hardy-like conditions
($i=1,2,\cdots,2n+3$) are
\begin{equation}\label{2class-J-odd-B}
  P(Z_{i}=1,Z_{j_1}=-1,Z_{j_2}=-1,\cdots,Z_{j_{2n+2}}=-1) = \frac{1}{2n+4},
\end{equation}
and
\numparts\label{2class-C-odd-B}
\begin{eqnarray}
  &P(X_{j_1}X_{j_2}\cdots X_{j_{2n+2}}=-1|Z_i=1)= 1,\label{2class-C-odd-B-1}\\
  &P(X_iX_{k_{2n+2}}=1|Z_{k_1}=Z_{k_2}=\cdots=Z_{k_{2n+1}}=-1)=1,\label{2class-C-odd-B-2}
\end{eqnarray}
\endnumparts
where $j_1\neq j_2\neq\cdots\neq j_{2n+2}\in\{1,2,\cdots,2n+3\}\backslash\{i\}$ and $k_1\neq k_2\neq\cdots\neq k_{2n+2}\in\{j_1,j_2,\cdots,j_{2n+2}\}$.
Note that equation \eref{2class-C-odd-B-2} contains $2n+1$ similar relations.

One can check that a total of $2n+4$ HLQP paradoxes can be induced. Combining them together will give rise to a DAVN proof of Bell nonlocality.

\end{document}